\documentclass{jpsj3}
\usepackage{txfonts}
\usepackage[dvipdfmx]{graphicx}

\title{Revisitation of the lock-in transition in one-dimensional charge-density waves}

\author{Katsuhiko Inagaki$^1$\thanks{kina@asahikawa-med.ac.jp}, Keiji Nakatsugawa$^2$, and Satoshi Tanda$^3$}

\inst{$^1$ Deparment of Physics, Asahikawa Medical University, Midorigaoka Higashi 2-Jo, Higashi 1-Chome, Asahikawa, 078-8510, Japan\\
$^2$ Research Center for Materials Nanoarchitectonics (MANA), National Institute for Materials Science (NIMS), Tsukuba 305-0044, Japan\\
$^3$ Division of Applied Physics, Graduate School of Engineering, Hokkaido University, Kita 13 Nishi 8, Kita-ku, Sapporo 060-8628, Japan}

\abst{
We studied the lock-in transition of charge-density waves of one-dimensional conductors. Though this phenomenon has been known for decades, there are still discrepancies between the theories and the experiments. We focused on the pioneering study of this phenomenon by McMillan and revisited, in particular, his numerical calculations. We first reproduced his results by providing unwritten information in the article. The obtained critical point was in good agreement with the previous study, as well as the analytical solution proposed later. 
We found a simple power-law behavior of the free energy in the critical regime, leading to the power-law divergence of the specific heat. We also introduced the concept of stiffness, equivalent to helicity modulus, which also showed a power-law behavior. These results allowed us to propose a new perspective on the lock-in transition.
}

\begin{document}
\maketitle

Since the first experimental result of the lock-in transition in charge-density waves (CDWs) was reported in 1975 \cite{Denoyer1975, Kagoshima1975, Comes1975}, it has been an attractive topic both theoretically and experimentally \cite{Bak1976, McMillan1976, Nakanishi1977, Roucau1980, Suits1980, Chen1981, Roucau1983, Fleming1984, Tanda1984, Tanda1985, Fleming1985, Fukuyama1985, Sagar2008, Inagaki2008,  Sakabe2017, Inagaki2018, Nakatsugawa2020}. 
The wavelength of a CDW $\lambda_\mathrm{CDW}$ is determined by the nesting vector connecting over the Fermi surfaces, independent of the lattice constants of the pristine lattice. Hence the ratio $M=\lambda_\mathrm{CDW}/a$ does not have to be an integer, where $a$ denotes any of the lattice constants.
When $M$ is close to an integer, the CDW may be locked by the pristine lattice at low temperatures so that $\lambda_\mathrm{CDW}$ shifts to an integral multiple of $a$. This phenomenon is called the lock-in transition.
One of the distinguished features in the lock-in transition is the emergence of solitons.  
Two pioneering papers argued this issue, one by Bak and Emery \cite{Bak1976}, and another by McMillan \cite{McMillan1976}. In both papers, the energy gain of the lock-in transition is attributed to the Umklapp process. A significant difference between them is the arrangement of solitons. According to Bak and Emery, a single soliton may exist as a solution of the sine-Gordon equation. On the other hand, McMillan's theory predicted a periodic lattice of solitons, namely, discommensurations.
Various experiments followed to show that discommensurations emerge in CDWs of quasi-two-dimensional (q2D) conductors \cite{Suits1980, Chen1981, Tanda1984, Tanda1985, Sakabe2017, Nakatsugawa2020, Nakatsugawa2024}. 
Extension of the theory to the q2D conductors was later performed and compared with the experiments \cite{Nakanishi1977, Nakatsugawa2020, Nakatsugawa2024}. 

In contrast to the q2d conductors, the lock-in transition of quasi-one-dimensional (q1D) conductors has not clearly been understood 
even though the first calculations actually treated the 1D case \cite{Bak1976, McMillan1976}. 
An analytic solution provided by Fukuyama and Takayama \cite{Fukuyama1985} reproduced the essence of McMillan's numerical result. We will see an overview of these theories as follows.
McMillan's theory on the lock-in transition of CDWs \cite{McMillan1976} concerned the 1D case. The position-dependent complex order parameter $\psi(r)$ develops by the Peierls transition. In general, the CDW wave vector $\vec{q}$ is incommensurate (IC), where the ratio $M$ is not an integer \cite{Gruner}. The appearing CDW is proportional to the real part of the order parameter, $\alpha(r) = \mathrm{Re}\{\psi(r)\}$. The free energy $F$ is written as:
\begin{eqnarray}
F &=& \int d^2r [a\alpha^2 -b \alpha^3 + c\alpha^4+ d|(\vec{q}\cdot\vec\nabla-iq^2)\psi|^2],\label{eq:Landau}
\end{eqnarray}
where $a$, $b$, $c$, and $d$ are periodic functions, such as
\begin{equation}
a= a_0 + a_1 \left( e^{i \vec{K}\cdot \vec{r}} + \mathrm{c.c.}\right),
\end{equation}
and $\vec{K}$ denotes the reciprocal-lattice vector. The complex conjugate represents
indeed $-\vec{K}$, because the CDW wave vector $\vec{q}$ lays parallel to $\vec{K}$  for a q1D system.
The Landau theory of phase transition determines
$a$ and $c$ for the system to undergo the Peierls transition \cite{McMillan1975}. The coefficient $b$ denotes the third-order Umklapp process responsible for the lock-in transition \cite{Lee1974}. 
Moreover, $d$ determines the elastic energy by spatial deformation. 
The third-order Umklapp process is included to treat the case $M=3$, 
typically observed in q2d conductors. Provided the CDW wave vector $q$ is close to commensurate, $\vec{q} \simeq \vec{K}/3$, and the order parameter modulated only by phase, $\psi(\vec{r})=\psi_0e^{-i\vec{K}\cdot \vec{r}/3}\phi(\vec{r})$ and $\phi(r)=e^{-i\theta(x)}$, where $\vec{s}=x = (-q+K/3)\vec{r}$, the following formula is obtained:
\begin{eqnarray}
f&=&\int dx [Y(1-\cos 3\theta ) +(d \theta /dx -1)^2].\label{eq:energy}
\end{eqnarray}
Here $f$ is the reduced free energy normalized by $F^0\beta$ \cite{Comment1}. By increasing $Y$, which depends on temperatures, the lock-in transition occurs at the critical point $Y_c$. Since $\theta(x)=0$ in commensurate CDW (C-CDW), eq. (\ref{eq:energy}) becomes $f=x_{max}$, where $x_{max}$ is the size of the system. For $Y > Y_c$, eq. (\ref{eq:energy}) gives $f/x_{max} >1$. 

McMillan assumed that the phase profile $\theta(x)$ was a superposition of a linear term and local modulation as ansatz:
\begin{eqnarray}
\theta(x)&=&\delta x +\sum_{n=1}^{N}A_n\sin(3n\delta x),\label{eq:ansatz}
\end{eqnarray}
where $\delta$ is the average slope of the phase, namely, normalized wave number, and $N$ is the number of Fourier terms. 
His strategy included two steps: First, minimizing $f$ with respect to $A_n$ while keeping $Y$ and $\delta$ fixed. Second, minimizing $f$ with respect to $\delta$ using the optimized $A_n$ to find the optimum wave number at a given temperature.
He found that the wave number $\delta$ continuously decreased to zero at $Y_c=1.2337$ and that the specific heat diverged at $Y_c$. Both are consistent if \textit{the transition is second-order}. Moreover, near $Y_c$, the wave profile should include a periodic lattice of phase jumps, or discommensurations. Each phase jump is accompanied by a fractional charge $2e/3$.

The first analytical approach to the lock-in transition was performed by Bak and Emery \cite{Bak1976}. They derived the sine-Gordon equation to describe this phenomenon and obtained a one-soliton solution. The soliton corresponds to the phase jump or a discommensuration in terms of McMillan's theory. However, there is an essential difference between their theories. McMillan predicted the emergence of a soliton lattice, while Bak and Emery proposed that a soliton emerges individually. They stated that interactions between the solitons might form a soliton lattice. A standard textbook follows the latter perspective on this issue \cite{Gruner}.

Later, Fukuyama and Takayama obtained a more general analytical solution that is in line with McMillan's way \cite{Fukuyama1985}. 
They started with the following formula:
\begin{eqnarray}
E &=& \int dx [A (\nabla \phi-q_0)^2 - B \cos \phi]. \label{eq:FT}
\end{eqnarray}
where $A$ and $B$ are constants, and $q_0$ is the wave number.
They derived that the lock-in transition occurs at $q_0=Q_c=(8B/A\pi^2)^{1/2}$.
They also found the $y$ $(=Aq_0^2/B)$ dependence of $q$. Near the transition, 
\begin{eqnarray}
\delta &\propto & [\ln (y-y_c)]^{-1} \label{eq:FT2},
\end{eqnarray}
where $y_c=1/Y_c$ corresponds to the critical point.
Its origin was not written, this behavior corresponds to the critical slowing down of the period of a pendulum, whose velocity is close to the branch point between oscillation and continuous rotation. 
Though they did not show the value $Y_c$ in McMillan's representation, one may calculate that $Y_c=\pi^2/8=1.233700550\ldots$, which agrees with the previous study.
The obtained wave profile was written in terms of Jacobi's elliptic functions.
They also exhibited that the wave profile includes discommensurations near the critical point. However, a quantitative comparison to McMillan's theory was not performed. The physical picture of the logarithmic behavior, shown in (\ref{eq:FT2}), was also not described. Hence, it remains necessary to evaluate McMillan's numerical calculations.

In this study, we focused on McMillan's numerical calculations. There are discrepancies between the theories and the experimental results, particularly for the q1D CDWs. McMillan predicted second-order transition, while the recent experimental results showed traces of first-order transition\cite{Sagar2008, Inagaki2008}. On the other hand, an unexpectedly long correlation length of q1D CDW in TaS$_3$ was reported \cite{Inagaki2010, Tsubota2012}. This might be explained by the existence of the discommensurations as described by McMillan, that is to say, not by a single soliton. Therefore, the periodicity of solitons should also be an important issue.

\begin{figure}
\center
\includegraphics[width=0.6\textwidth]{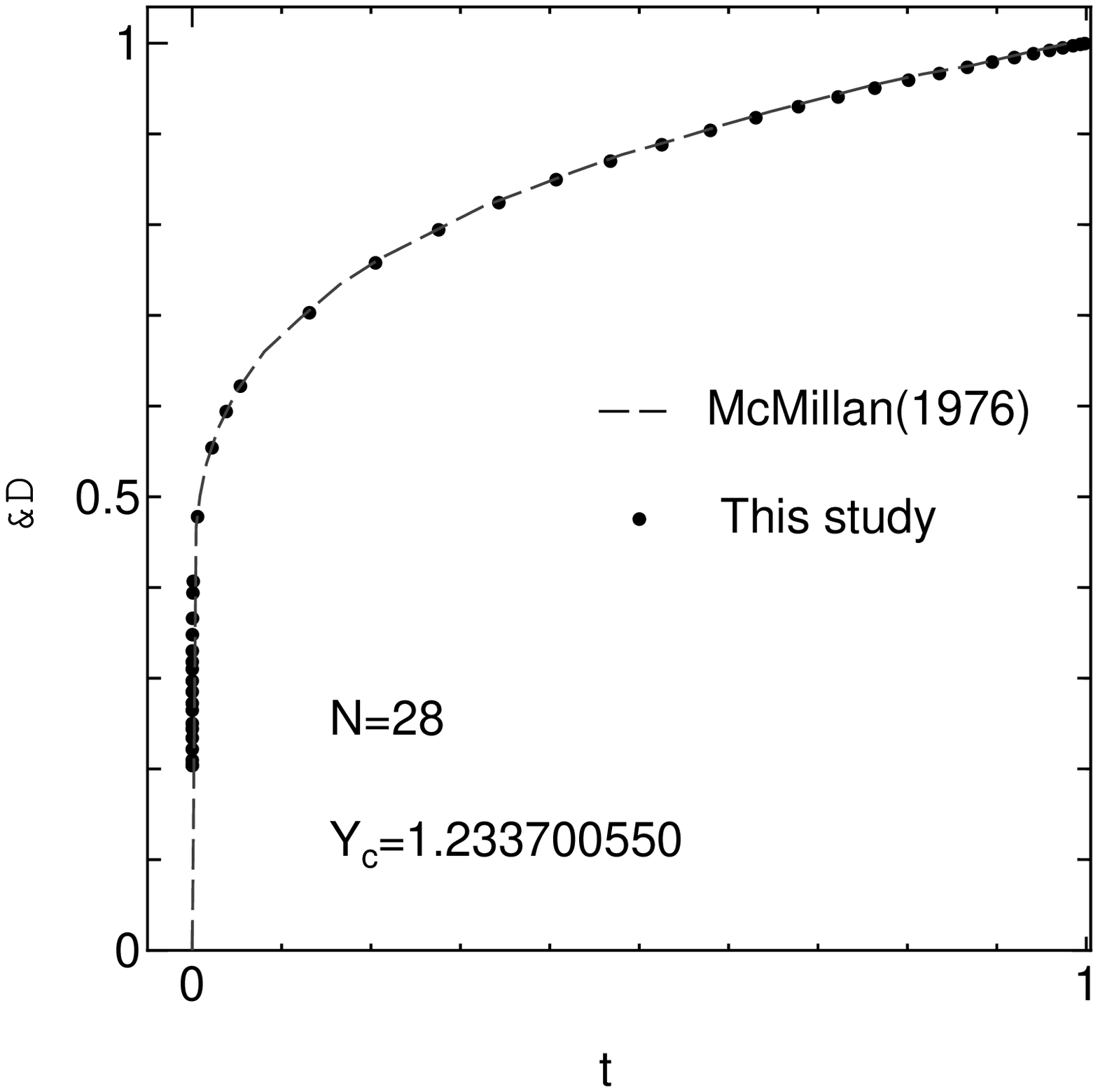}
\vspace*{3cm}
\caption{Normalized wave number $\delta$ a function of the reduced temperature $t$. The solid circles represent the result of this study. The broken line corresponds to Fig. 1 of Ref. \citen{McMillan1976}.}\label{fig:wavenumber}
\end{figure}

We performed numerical calculations based on McMillan's free energy, shown in eq. (\ref{eq:energy}).
Since the description of numerical calculation was insufficient in the previous study, we tested various algorithms.
As a result of the preliminary test, we used the Nelder-Mead method, Simpson's rule, and golden section search for optimization of the Fourier coefficients, numerical quadrature, and final adjustment of wave number, respectively. All of them were modified from those written in the textbook of numerical calculations \cite{NR}.  
In all calculations, we used extended double-precision variables, namely, 80-bit-long floating decimals.
We obtained 28 terms of Fourier coefficients for each calculation. It is known that the expected error is $\varepsilon^2$ in the 80-bit-long floating decimals, where $\varepsilon$ is the machine epsilon. Therefore, we set tolerance in the order of $10^{-10}$ in optimization, and $10^{-12}$ for quadrature. 
Figure \ref{fig:wavenumber} shows the normalized wave number $\delta$ as a function of the reduced temperature $t=(T-T_{CI})/(T_{IN}-T_{CI})=(Y_c^2-Y^2)/Y_c^2$, where $T_{CI}$ and $T_{IN}$ are the lock-in temperature and the Peierls temperature, respectively. The dashed line shows McMillan's result (Fig. 1 in Ref. [\citen{McMillan1976}]) and the solid circles represent the result of this study, showing that our numerical calculations successfully reproduced the previous one.
 We obtained the critical point $Y_c=1.233700550$, which agrees with the theory by $10^{-10}$ accuracy. It should be remembered that the previous study provided the same value to the $10^{-5}$ order, $Y_c=1.2337$. The difference in the precision plays an important role as we show below.

\begin{figure}
\center
\includegraphics[width=0.6\textwidth]{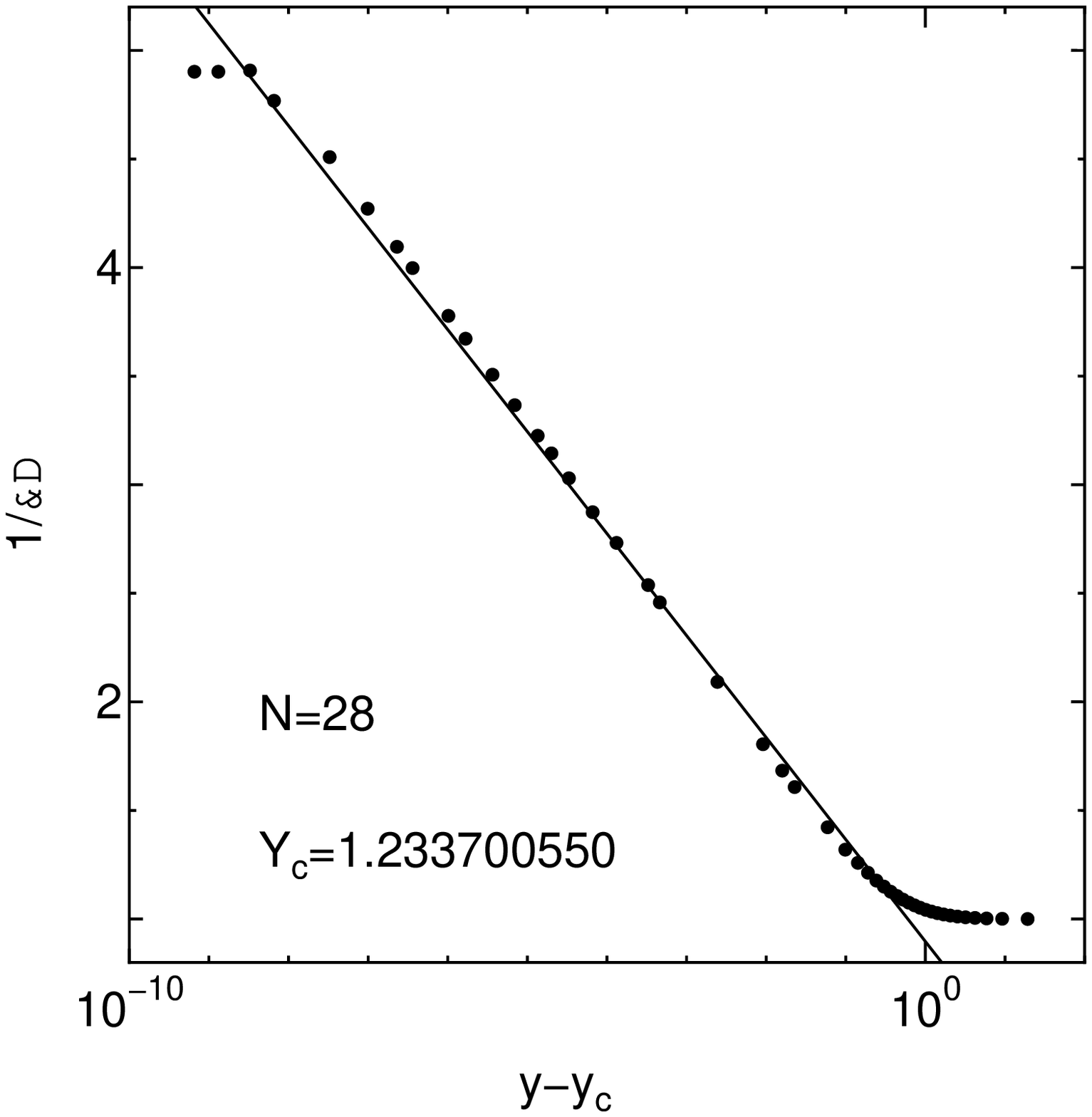}
\vspace*{3cm}
\caption{Inverse of the normalized wave number $1/\delta$ as a function of $y-y_c$.  The solid line is the guide to an eye.}\label{fig:log}
\end{figure}

An increase in precision allows us to understand critical behavior in the neighborhood of the critical point $Y_c$. Some empirical formulae were obtained by fitting the results in the previous study, as shown in eqs. (11), (12), and (13) of Ref. [\citen{McMillan1976}], for the energy gain, reduced wave number, and specific heat, respectively. It should be noted that these emperical formulae exclude near the critical point. In contrast, our study provides 10-digit precision for $Y_c$. First, the reduced wave number $\delta$ is plotted in Fig. \ref{fig:log}, $1/\delta$ as a function of $y-y_c$ to compare the analytic solution (\ref{eq:FT2}).
The straight line on these axes shows the predicted logarithmic behavior. As shown in Fig. \ref{fig:log}, our numerical results fell to a straight line in the wide range of $y-y_c$. Several points near $y_c$ do not seem to change probably because of the calculation limit. 
Though the empirical fitting curve by McMillan, $\delta(t)=4.62/[4.61+ \ln(1/t)]$, may be a good approximation for $0.005 < t < 0.5$, we compared the analytical and numerical result for the first time and obtained a good agreement in the neighborhood of the critical point, $10^{-9} <t < 0.1 $.

\begin{figure}
\center
\includegraphics[width=0.6\textwidth]{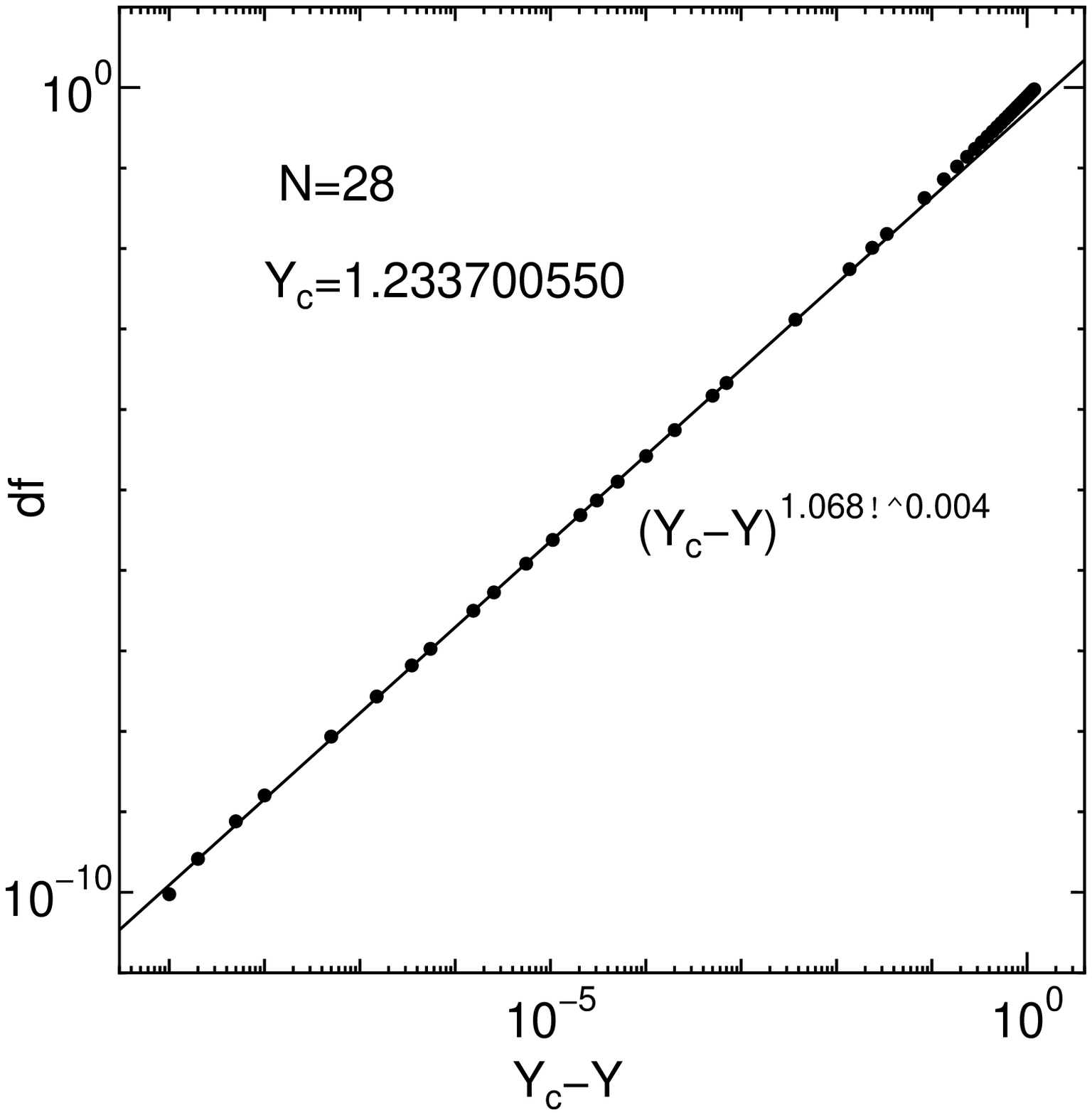}
\vspace*{3cm}
\caption{Log-log plot of the energy as a function of $Y_c-Y$. The solid line is the least-square fit of the data.}\label{fig:energy}
\end{figure}

One may be interested in whether the lock-in transition is truly second-order or not. In the previous study, McMillan proposed that the specific heat $C_v$ obeys $C_v=F^o\beta1.35/t \ln(7.4/t)$, implying second-order transition. We reconsidered this formula based on our numerical results. Figure \ref{fig:energy} shows the log-log plot of the energy gain $df=f-1$ as a function of $Y_c-Y$. The solid line is the least-square fit of the 20 points near $Y_c$. The energy obeys a simple power law $(Y_c-Y)^{1.068\pm0.004}$, from which we obtained
\begin{eqnarray} 
C_v &=& F^o\beta0.0218t^{-0.932}. \label{eq:Cv}
\end{eqnarray}
Even though the obtained formula differs, our results also support that the lock-in transition described with McMillan's free energy is second-order. 
Importantly, the power-law behavior (\ref{eq:Cv}) implies the lock-in transition in 1D-CDW should be treated in terms of critical phenomena. 
The origin of the power-law index -0.932 will be an issue to be solved in future studies.

Now let us discuss the physical pictures of the lock-in transition, by considering energies of I- and C-CDWs. 
Since we obtained the energy minima by sweeping the initial $\delta$, we also possess the energy profile as a function of $\delta$. 
No local minima were found around the true minimum for all $Y$. This is consistent with the picture of second-order transition. Moreover, this result suggests that the solution of eq. (\ref{eq:energy}) is rigorously determined for a given $Y$ and that the solution has a perfect periodicity with discommensurations. This also means the spacing between discommensurations becomes a constant, as well as $\delta$, for a given $Y$. 
In contrast, since the energies between I- and C-CDWs become equal at $Y_c$, solitons can be created and annihilated spontaneously. Instead of the discommensurations (soliton lattice) in I-CDW, there are solitons at random spacing, leading to the absence of periodicity in C-CDW. 

\begin{figure}
\center
\includegraphics[width=0.6\textwidth]{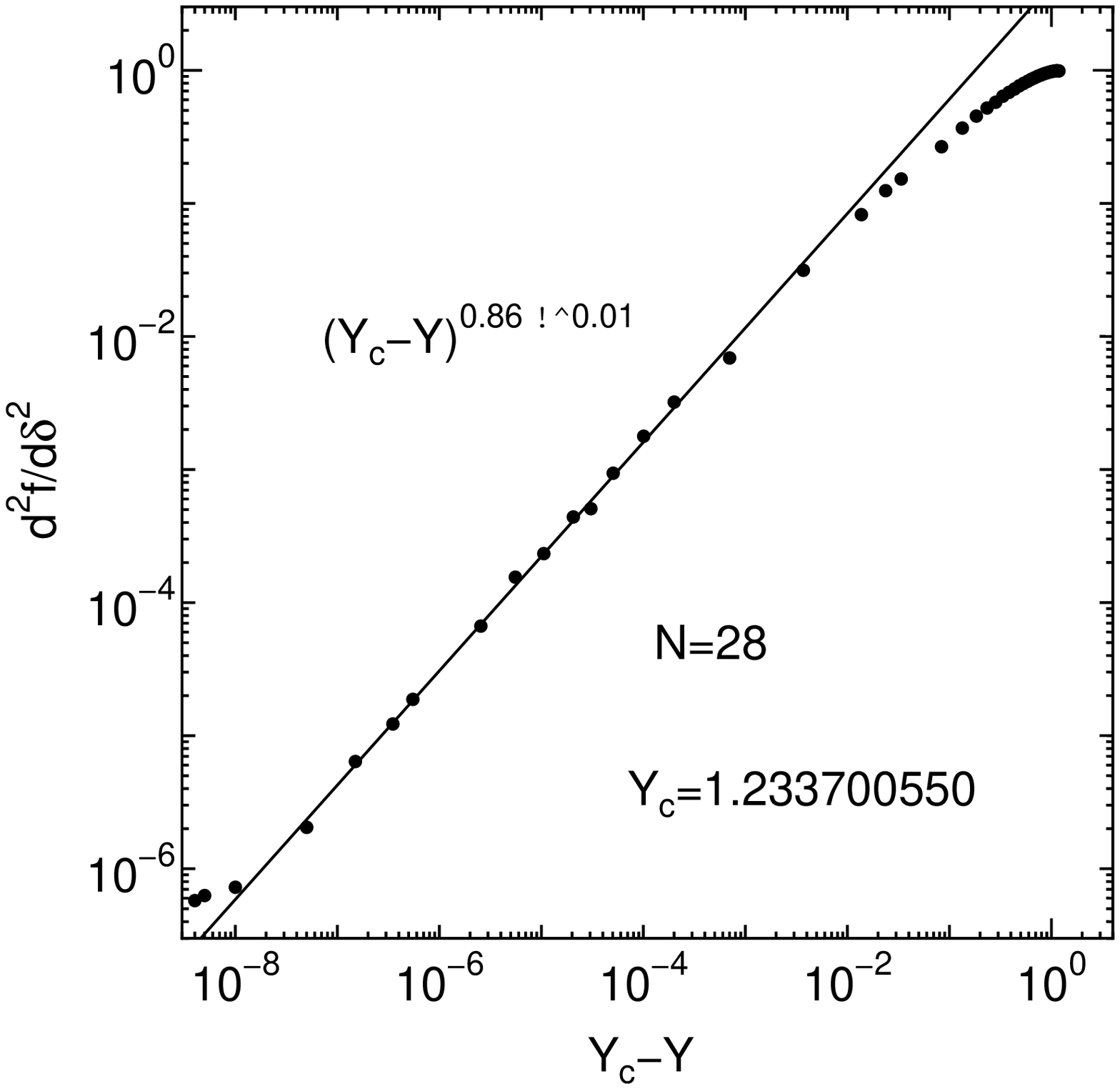}
\vspace*{3cm}
\caption{Log-log plot of $d^2f/d\delta^2$ as a function of $Y_c-Y$. The solid line is the least-square fit of the data.}\label{fig:stiffness}
\end{figure}

Moreover, the obtained energy profile allows us to provide the ``stiffness'' $d^2f/d\delta^2$ as a function of $Y_c -Y$ in Fig. \ref{fig:stiffness}. The vanishing stiffness implies the system became softened for $Y \to Y_c$. Importantly, the straight line in the log-log plot exhibits a power-law behavior near $Y_c$. The critical exponent was obtained as $0.86\pm0.01$ from the least-square fitting. The stiffness corresponds to the \textit{helicity modulus} in the context of phase transition \cite{Fisher1973}. A small change in $\delta$ induces a change in the system size $2\pi/3\delta$. If the system size is fixed, it is equivalent to the phase change at the end, which is in agreement with the definition of the helicity modulus. 

Therefore, Fig. \ref{fig:stiffness} proposes a new perspective on the lock-in transition. 
For the larger $Y$, the stiffness gets smaller. When $Y$ is close to $Y_c$, discommensurations, namely, a soliton lattice evolves and an I-CDW is cut into pieces of C-CDWs between the solitons. Since the stiffness is still finite, the system can be treated as a continuum. For $Y \to Y_c$, both the specific heat and the spacing between solitons diverge, described by the simple power-laws. Finally, at $Y=Y_c$, the system becomes C-CDW, where the stiffness vanishes. The system loses its continuity due to randomly created solitions.
That is to say, in contrast to the intuition, I-CDW should be \textit{ordered}, whereas C-CDW \textit{disordered}. 
This agrees with the picture we mentioned for Fig. \ref{fig:energy}.

In summary, we performed numerical calculations on the lock-in transition in 1D CDW, based on McMillan's free energy. 
Our results reproduced the previous study. 
By 10-digit precision calculations, we demonstrated consistency between the numerical and analytical results for the first time in the lock-in transition of 1D CDW. We found a simple power-law behavior of the free energy in the critical regime, leading to the power-law divergence of the specific heat. We also introduced the concept of stiffness, equivalent to helicity modulus, which also showed a power-law behavior. These results allowed us to propose a new perspective on the lock-in transition of 1D CDW.

\newpage

\newpage

\newpage

\newpage


\begin{thebibliography}{99}
\bibitem{Comment1}
In 2d CDWs, the coefficients are
 $F^0=a_0^2[3c_0(q-K/3)^2]^{-1}$, $\beta=e_0q^2\psi_0(q-K/3)^2/F^0$, and $Y=b_1\psi_0^2/[4e_0q^2(q-K/3)^2]$.
\end{thebibliography}
\end{document}